# On homogeneous statistical distributions exoplanets for their dynamic parameters

B. R. Mushailov, L .M. Ivanovskaya, V. S. Teplitskaya
Sternberg Astronomical Institute, Moscow State University Universitetsky pr., 13, Moscow 119992, Russia. *E-mail: brm@sai.msu.ru, ilm1655@yandex.ru VeraTeplic@yandex.ru*

**Abstract**
Correct distributions of extrasolar systems for their orbital parameters (semi-major axes, period, eccentricity) and physical characteristics (mass, spectral type of parent star) are received. Orbital resonances in extrasolar systems are considered. It is shown, that the account of more thin effects, including with use of wavelet methods, in obviously incorrectly reduced distributions it is not justified, to what the homogeneous statistical distributions for dynamic parameters of exoplanets, received in the present work, testify.

## Introduction

During dynamic evolution extrasolar planets pass a number of regular stages which have universal character. Revealing of these general laws, certainly, will promote better understanding of an origin and evolution of planetary systems.

At October, 2009 outside Solar system it was discovered about 400 candidates for major planets (extrasolar planets). A large number of extrasolar planets comparable in mass to Jupiter, but apparently this is a consequence of the selective effect of detection - giant planets easier to detect. In connection with the improvement of the observation base, development of new methods of detection, the number of candidates for extrasolar planets is steadily increasing. The most effective are the methods of searching candidates for extrasolar planets:

- The measuring of radial velocity. The star having a planet, tests fluctuations of velocity “ to us - from us ” which can be measured by observing the Doppler shift of the spectrum of the star.

- The photometric method, is connected with an opportunity of passage of a planet on a background of a star. The planet eclipses a part of a surface, and brightness of a star decreases.

- Astrometric method. It is based on taking into account of gravitational influence of planets on a star. If very precisely to measure a trajectory of a star it is possible to see its easy tortuosity caused by gravitation of planets.

- Microlensing. When one star passes another on a background, light of a distant star is bent by gravitation near star and its brightness varies. If the nearest star has planets it will affect a curve of change of brightness.

- A number of the methods based on obtaining of the direct image

- In addition, the method for ultra precise photometric monitoring of stars and method for a method of measurement of a curve of shine, but not due to transit of a planet, and in view of tidal influence from a planet (planets) (*Gerasimov I.A. et al.*, 2003) is developed also.

Currently, there are six specialized discusses space projects to find extrasolar planets (PEGASE, New Worlds Mission, IRSI/ DARWIN (ESA), Space Interferometry Mission SIM, Terrestrial Planet Finder TPF, Kepler (NASA), GAIA (ESA)) are discussed.

## The statistical distribution of exoplanets.

Till now the conventional and consistent theory of formation of planetary systems and multiple stellarr systems is not created. It is considered, that stars with rather small mass, located more to the right of the main sequence, could be members of multiple stellar systems, and that the most part of these dynamic systems has broken up under action of tidal forces. But due to capture interstellar gas-dust matters, these stars were capable to form the planetary systems. On the other hand, stars can lose the planets being far from them and having with a central star weak gravitational connection. Except for major planets, around of stars can exist a minor planets (asteroids, comets, etc.). Configuration of large and minor planets of the solar system suggests that their dynamic evolution is strongly influenced by the effects of orbital resonance interactions. In some cases resonant effects can lead to steady orbital movements. Consequently, there are strong grounds for believing that orbital resonances should be widely disseminated not only in our solar system, but also in other planetary systems, which is confirmed by the results shown in Figure 1.

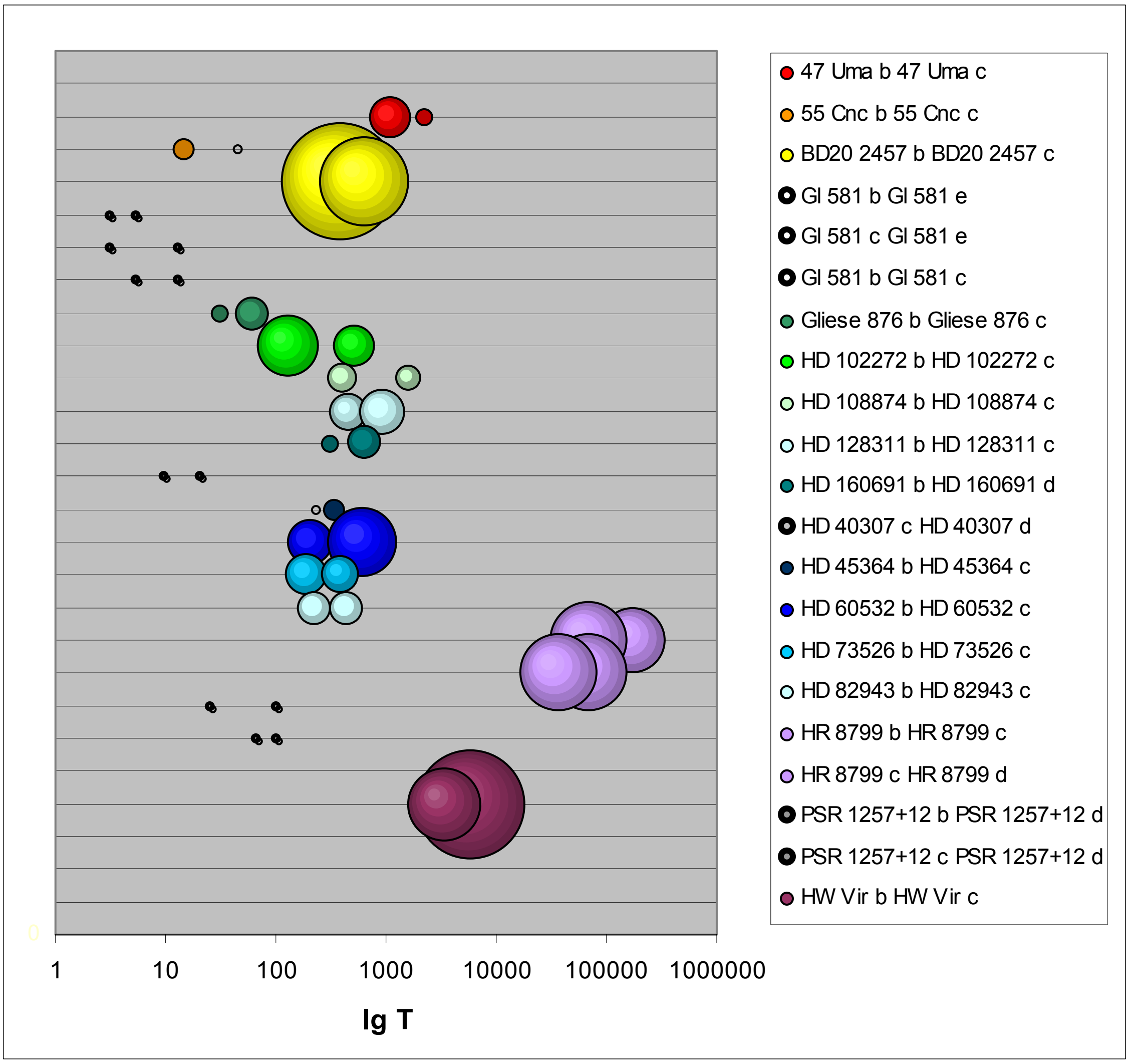

Fig. 1. The orbital periods of 17 alleged resonance exoplanets. We give the size of extrasolar planets in proportion to their masses, with the exception of three planetary systems, which are displayed as a points with shadow, in view of the smallness of their masses.

In Solar system, as a rule, dynamic parameters are measured in corresponding scale units of Earth and Sun. For example, to unit of length scale is accepted astronomical unit (distance from Sun to Earth), and the mass of celestial bodies are measured in solar masses. Self-consistent normalization should be applied for exoplanets, for "Sun-Earth" a-priori parameters are not adapted (internally consistent) for arbitrary exoplanets. In addition, the statistical distribution of planetary systems in the absolute values of their parameters, of course, ignores their hierarchy - the dominant role of the central star at evolution of planetary system within the limits of the *n*-body problem, instead of all statistical sample of planets. Therefore, to determine the true (not diluted incorrect normalization) the patterns of evolution of extrasolar planets and the solar system, it is advisable to normalize the dynamic parameters of the values associated with the study of planetary system that has been done in this paper.

Figures 2-4 show the comparative distribution histogram of exoplanets for semi-major axes, masses, period (not adapted – according to http://exoplanet.eu/catalog-all.php and self-consistent units) for those modeling exoplanets, at which own parameters are known, and also the distribution of exoplanets for eccentricities. There are black columns for distributions in not adapted system of units and white columns for self-consistent system of units.

Transition to the self-consistent parameters is not only linear scale shift as for each planetary system it is entered own normalization. Consequently, found differences in the distributions presented below in the self-consistent and not adapted to the parameters not only due to the difference in scale sampling interval histograms, and are reflection of distinctions of directly presented distributions. Relevance of similar histograms with different intervals of samples confirms the reliability of the results presented below. From the distribution of extrasolar planets for semi-major axes of the planets that the distribution of a self-consistent system of units is more obviously the expressed maximum at the beginning, but decreases rapidly in comparison with the distribution in non-adapted system of units.

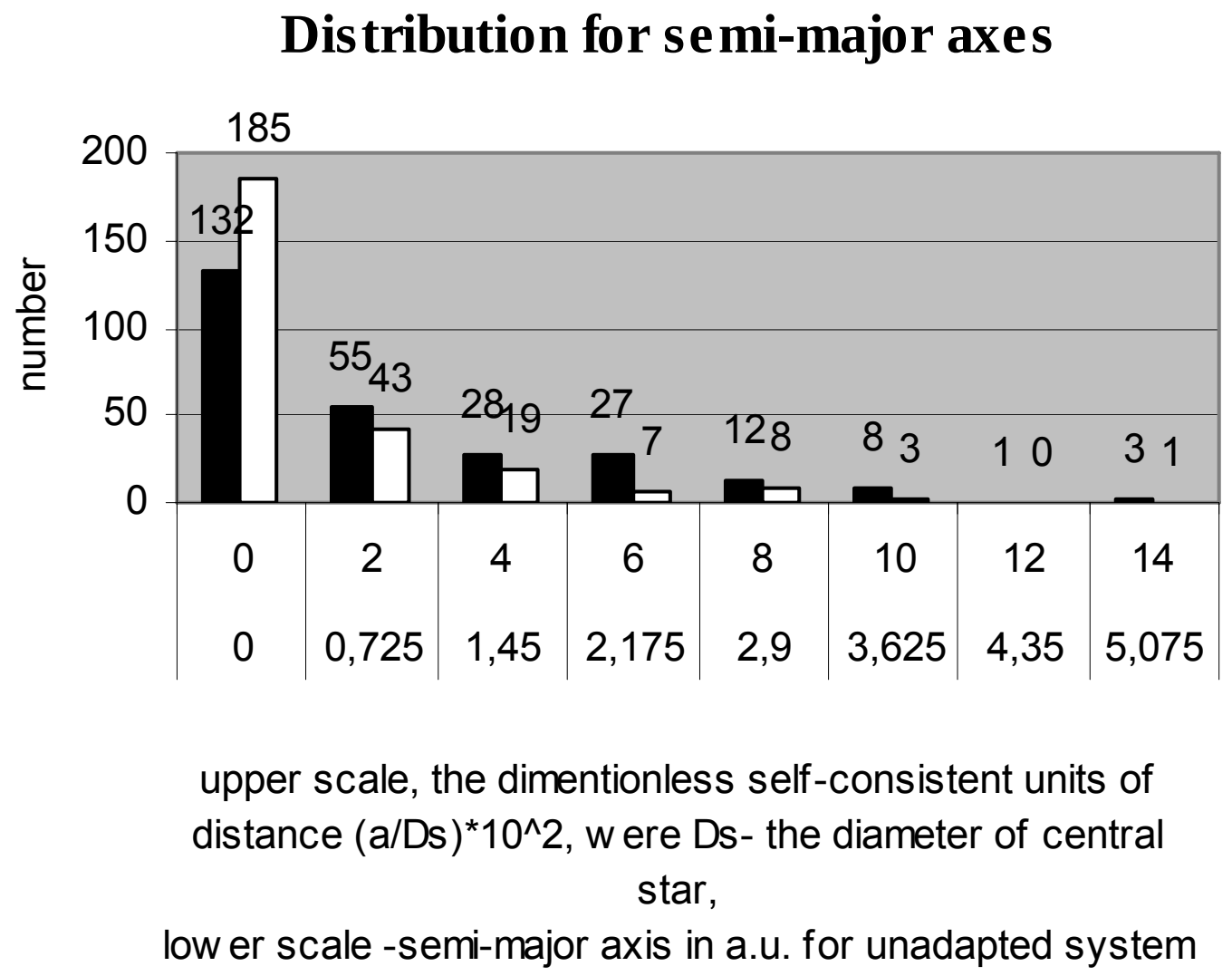

Fig. 2. The histogram distribution of extrasolar planets in semi-major axes . The correlation coefficient values in the self-consistent and not adapted systems of units is equal to 0.821.

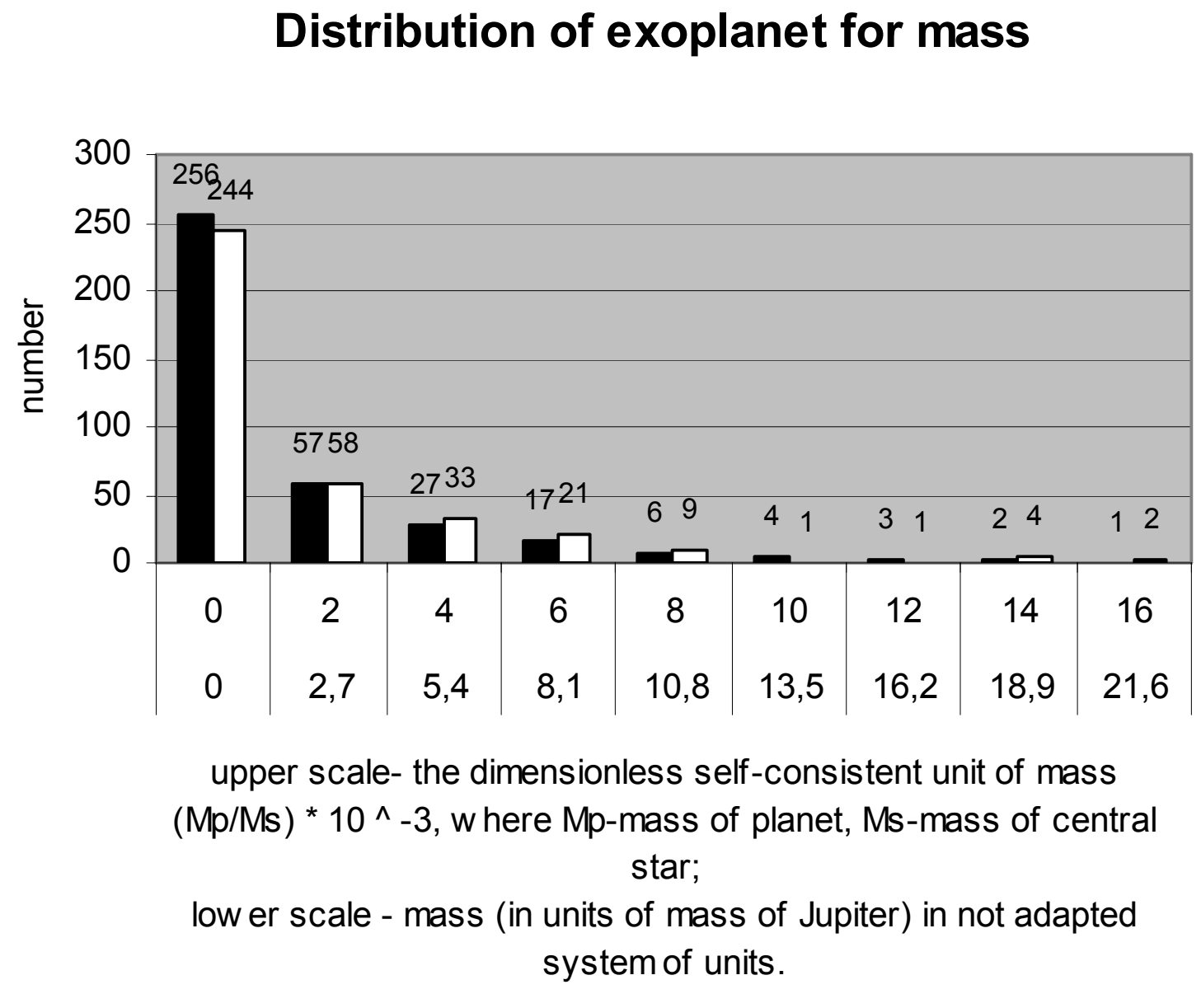

Fig. 3. The histogram distribution of extrasolar planets by the masses. The correlation coefficient values in the self-consistent and not adapted systems of units for the first five intervals equal to 0.894.

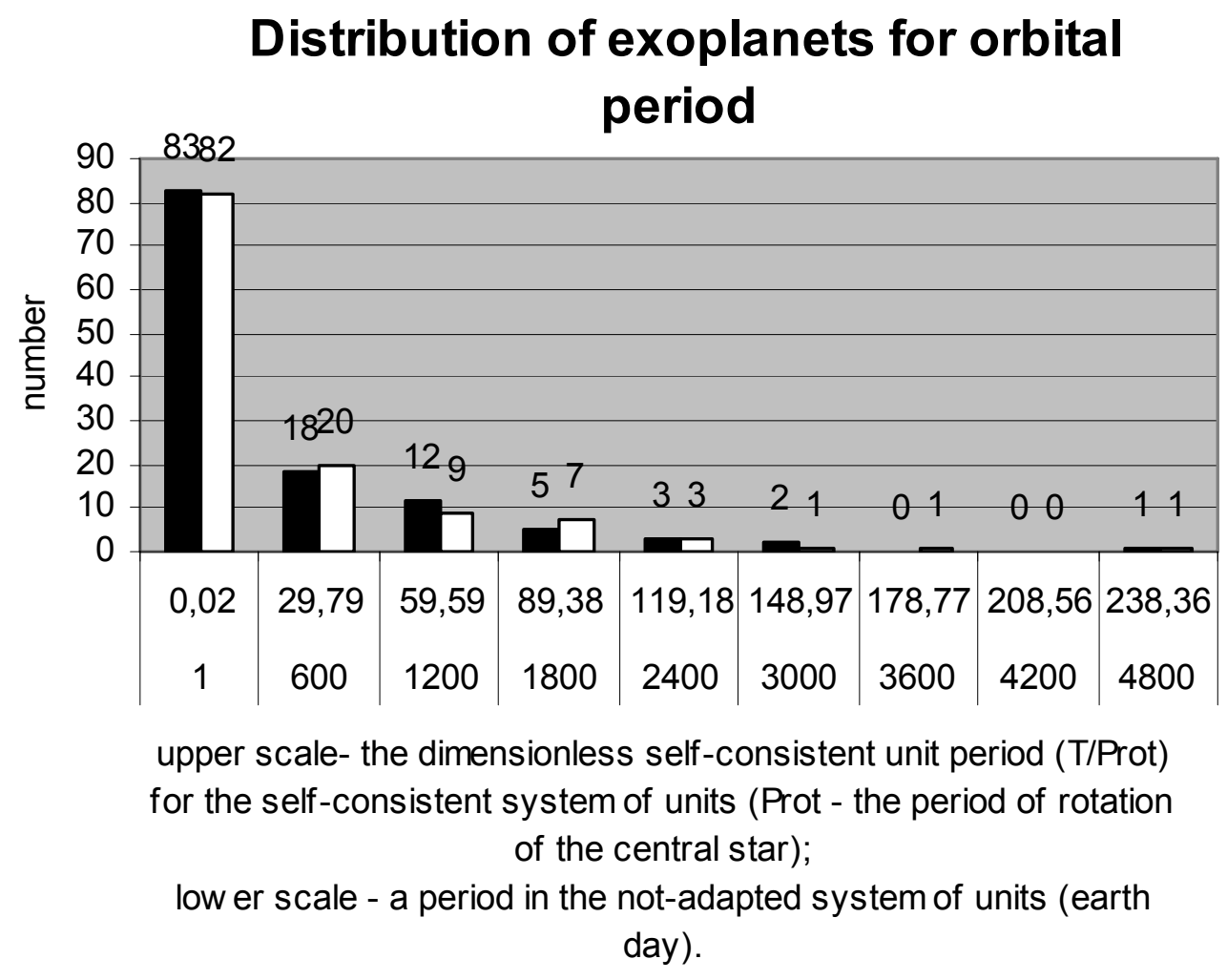

Fig. 4. The histogram distribution of extrasolar planets for orbital period. The correlation coefficient values in the self-consistent and not adapted systems of units is equal to 0.752 **.**

As follows from histograms of distributions of semi-major axis, the masses of the planets and their orbital periods, distributions in not adapted and self-consistent systems of units differ also it is not connected with distinction of volumes compared samples. The calculated corresponding correlation coefficients testify to nonlinear change of histograms at transition from not adapted units to self-consistent.

It is expressed that the differences in distributions in not adapted and self-consistent units for the fixed stellar spectral types (Fig 5.) should be less. On number of candidates for extrasolar planets, spectral type *G*, to which our Sun belongs, also is allocated.

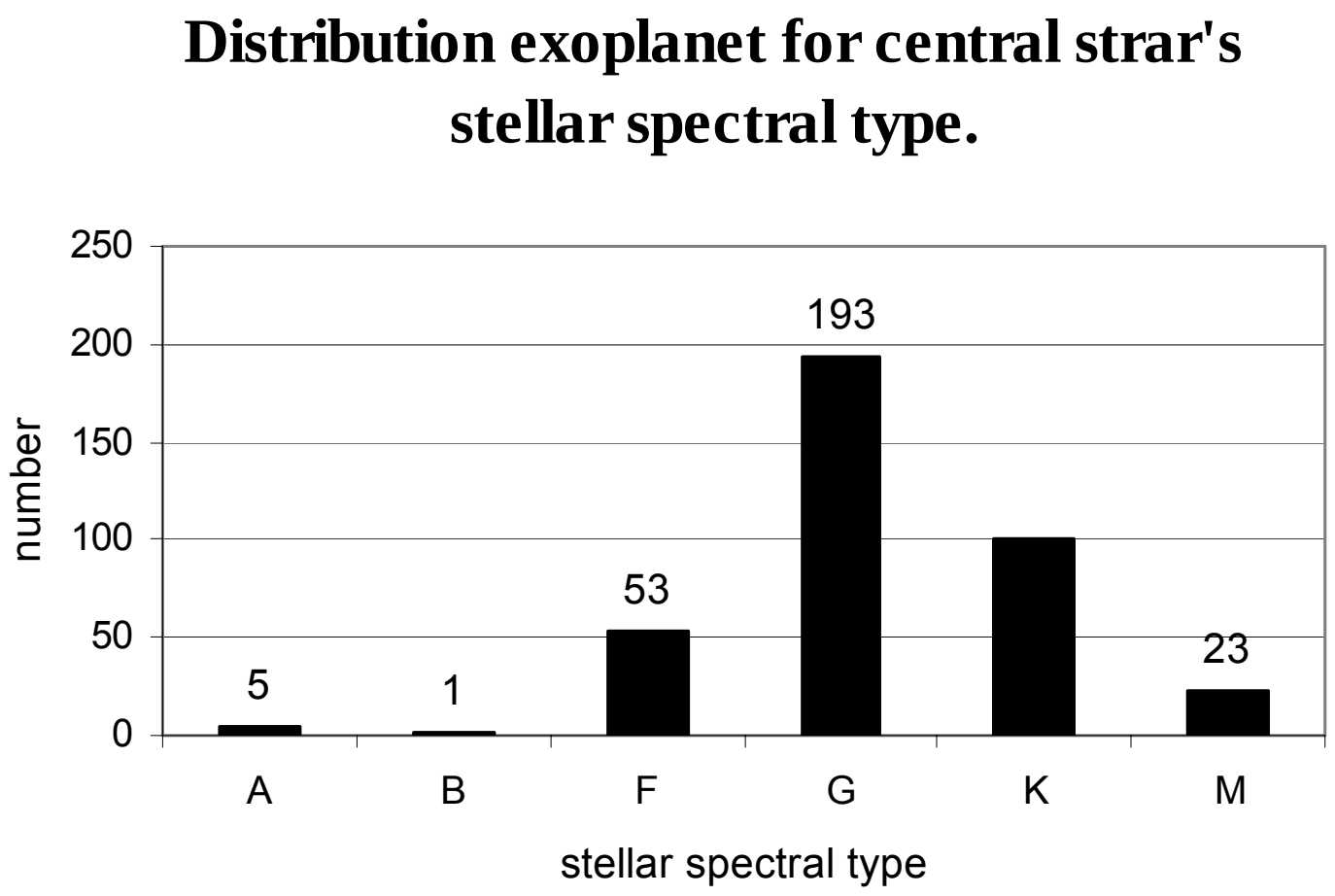

Figure 5. The histogram distribution of extrasolar planets by the stellar spectral type.

The adapted distributions considered exoplanets for semi-major axes almost for all spectral classes of the central stars are generally exponential in nature and monotonous.

Figures 6-14 are histograms of the self-consistent distributions of extrasolar planets for semi-major axes, the masses and orbital periods of planets for of spectral classes F, G, K. Classes A, B, M are excluded due to the small sample size.

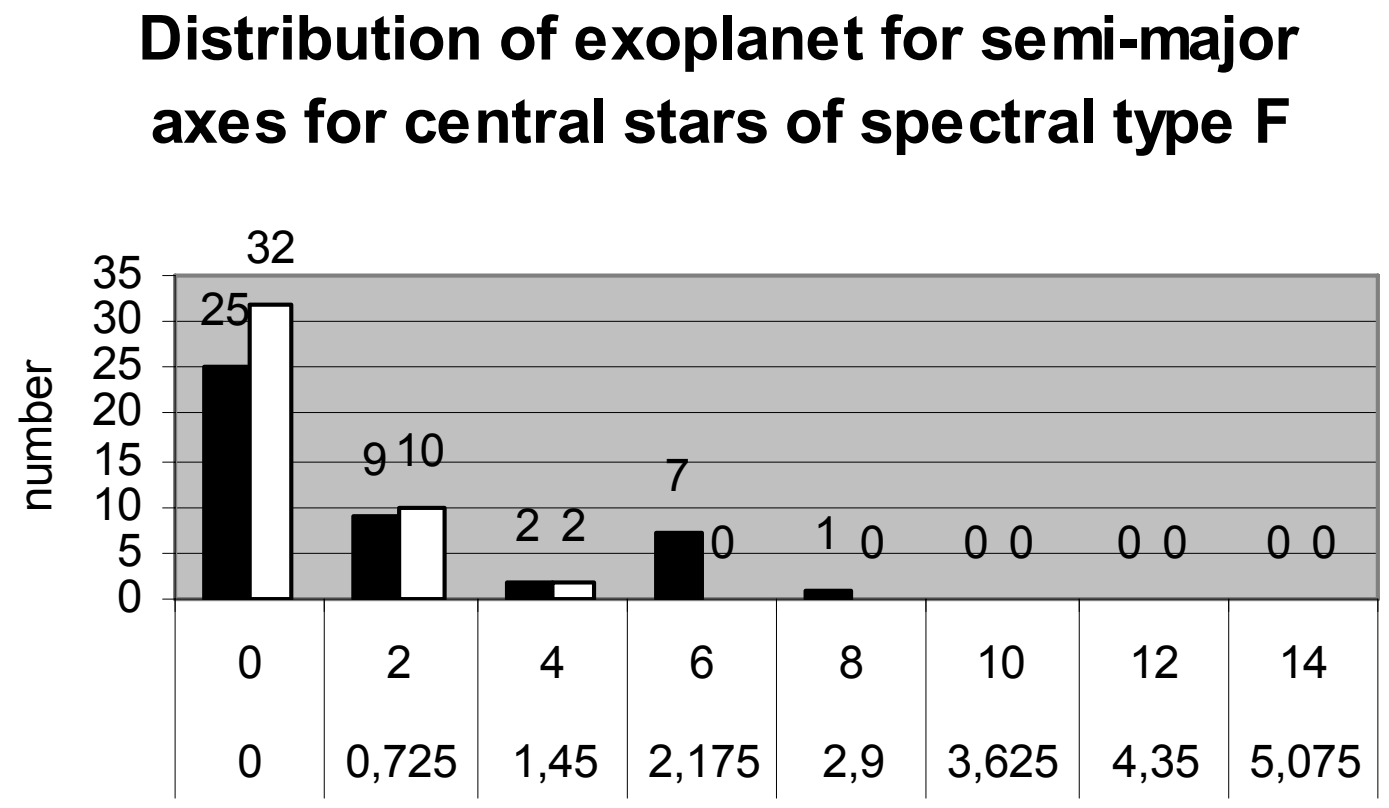

Fig. 6. The histogram distribution of exoplanets for semi-major axes of the planets for the stars of stellar spectral type F. The correlation coefficient values in the self-consistent and not adapted systems of units is equal to 0.957.

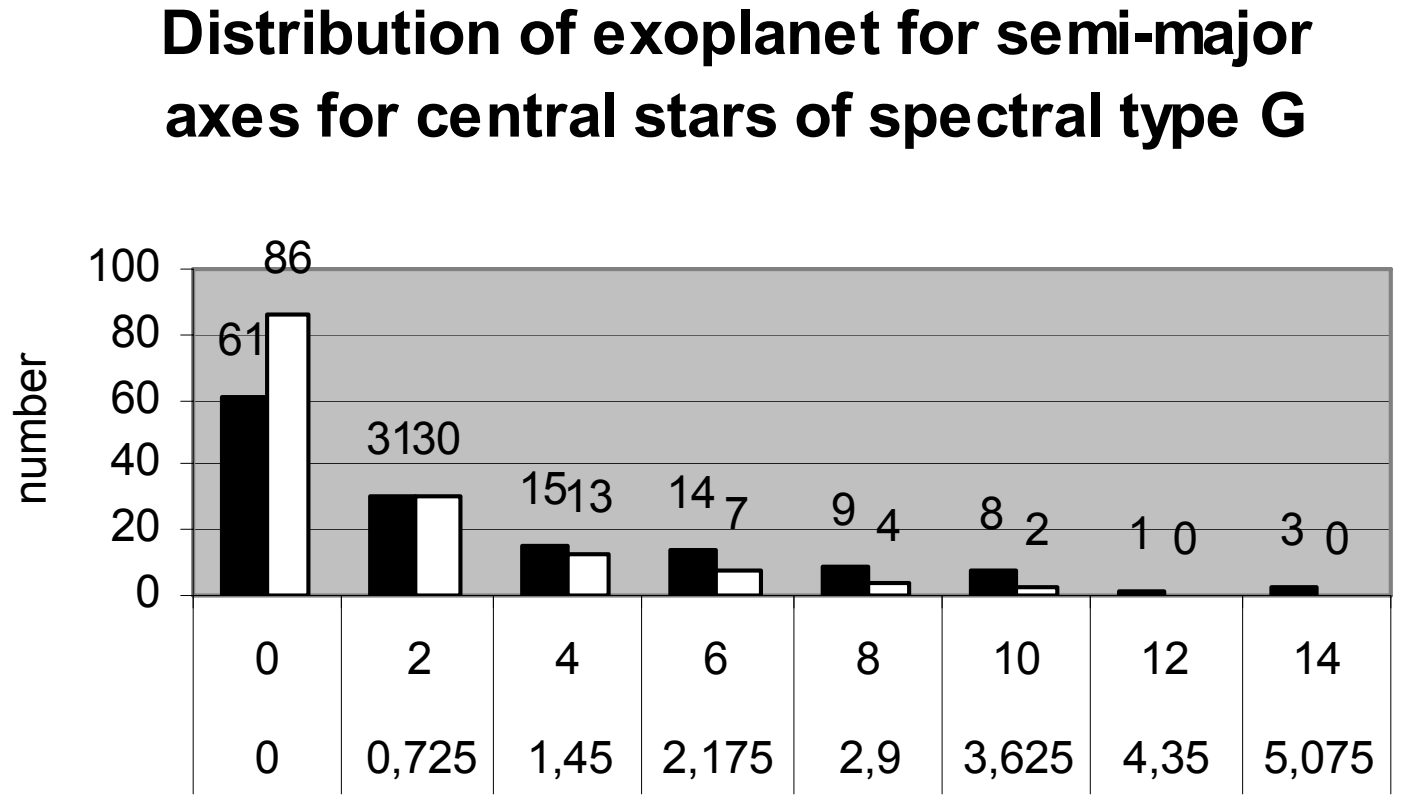

Figure 7. The histogram distribution of exoplanets for semi-major axes of the planets for the stars of stellar spectral type G. The correlation coefficient values in the self-consistent and not adapted systems of units is equal to 0.922 .

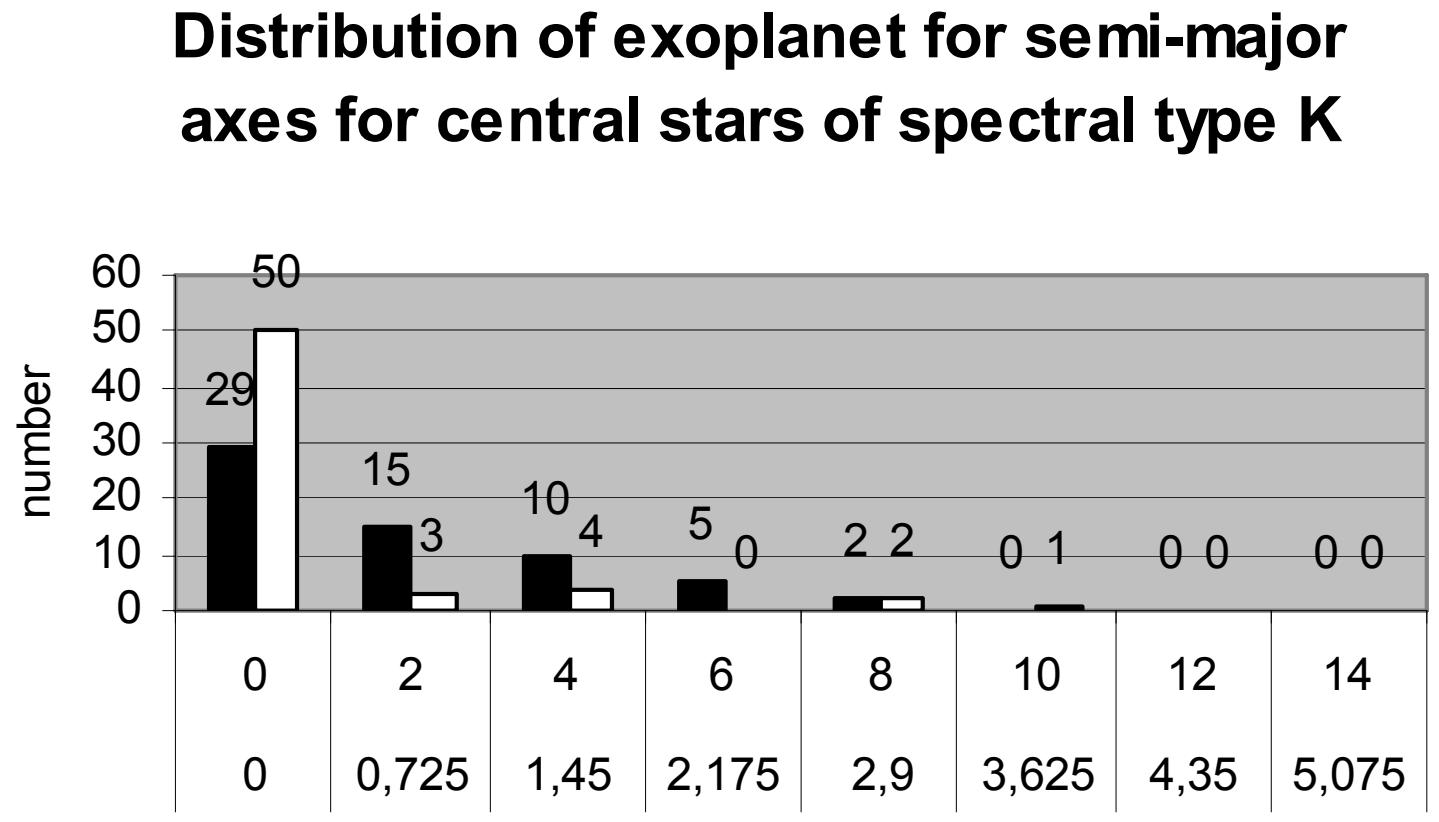

Fig. 8. The histogram distribution of exoplanets for semi-major axes of the planets for the stars of stellar spectral type K. The correlation coefficient values in the self-consistent and not adapted systems of units is equal to 0.696.

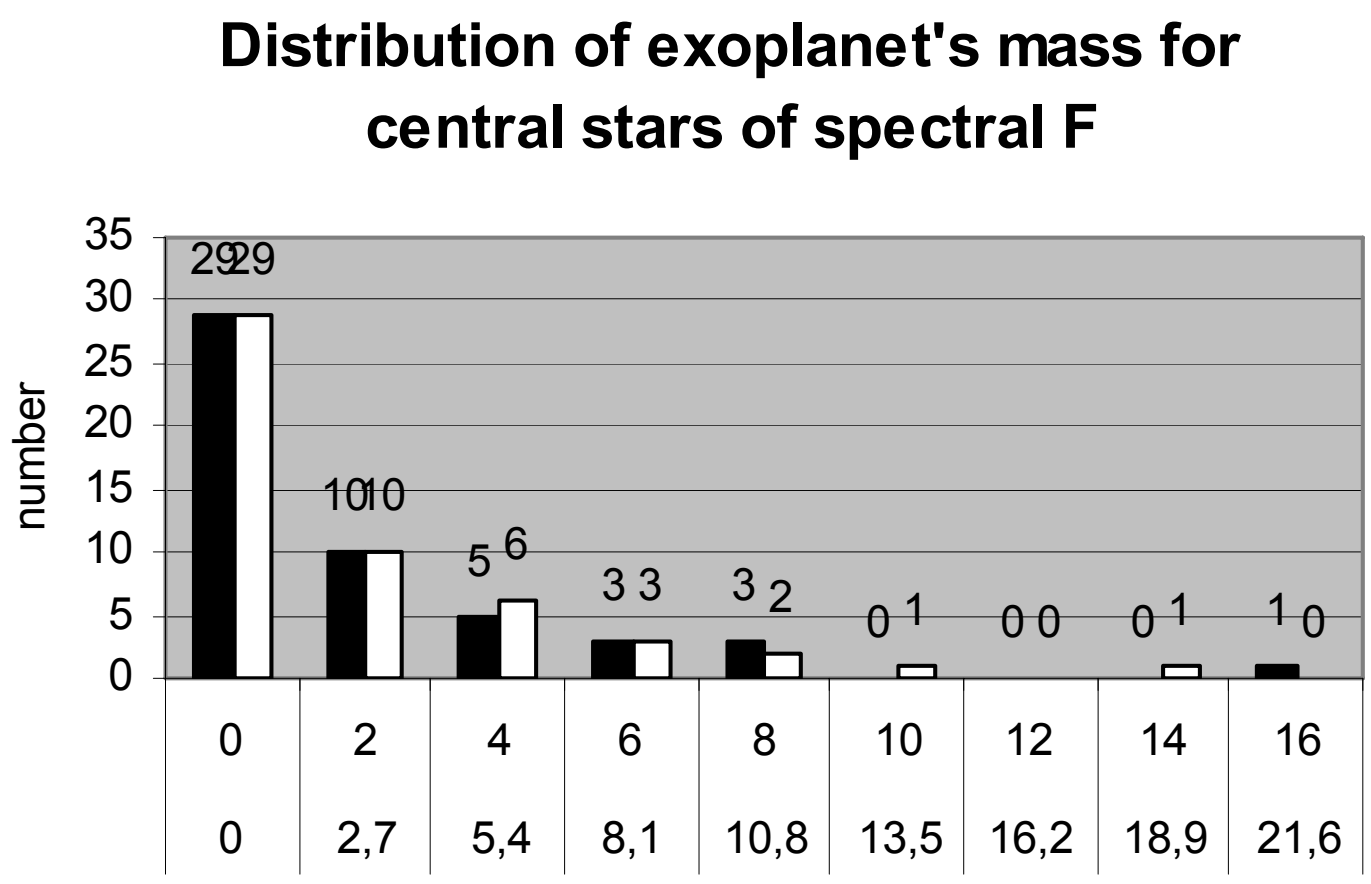

Fig. 9. Histogram distributions of extrasolar planets by the masses for stars of stellar spectral type F. The correlation coefficient values in the self-consistent and not adapted systems of units is equal to 0.978.

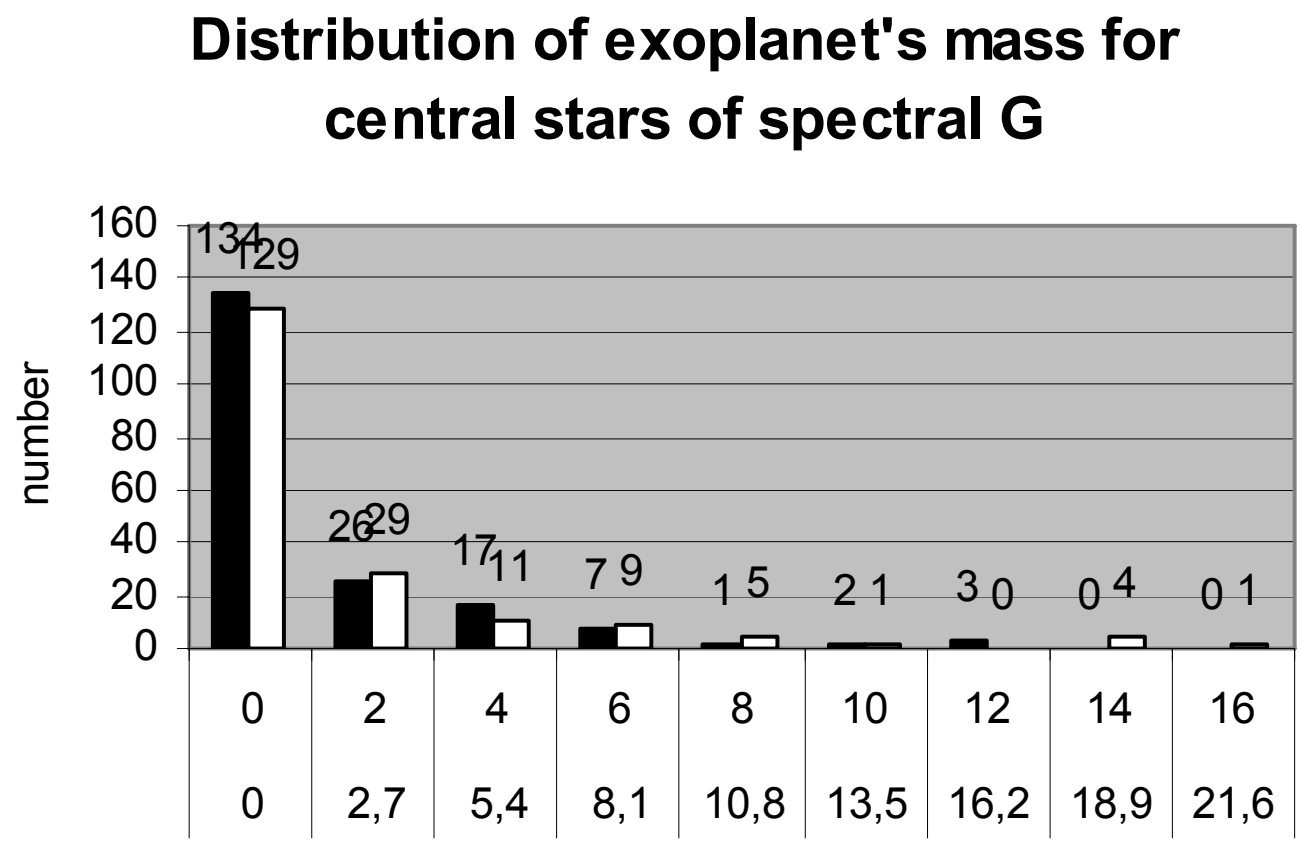

Fig. 10. The histogram distributions of extrasolar planets by the masses for stars of stellar spectral type G. The correlation coefficient values in the self-consistent and not adapted systems of units is equal to 0.975.

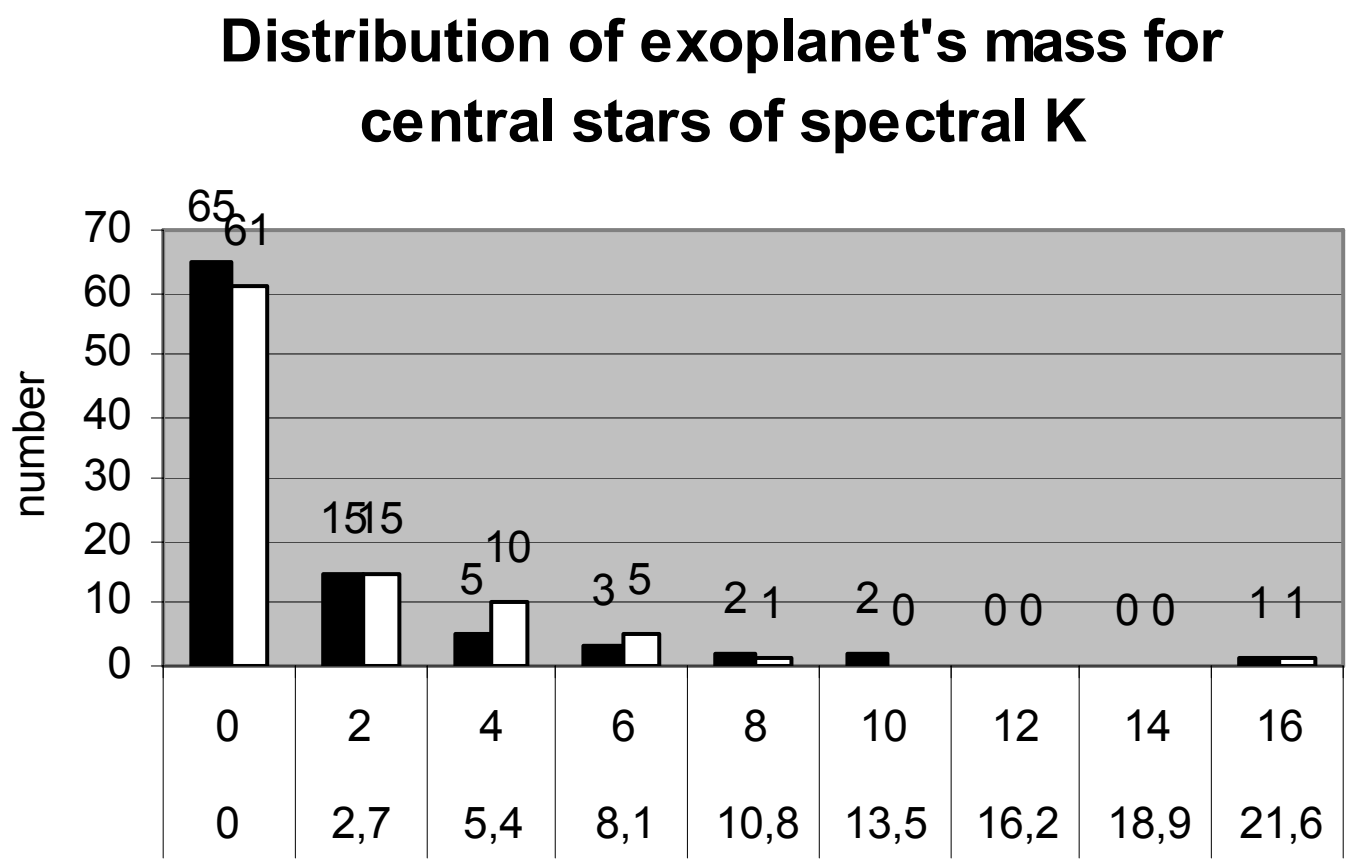

Fig. 11. The histogram distributions of extrasolar planets by the masses for stars of stellar spectral type K. The correlation coefficient values in the self-consistent and not adapted systems of units is equal to 0.791.

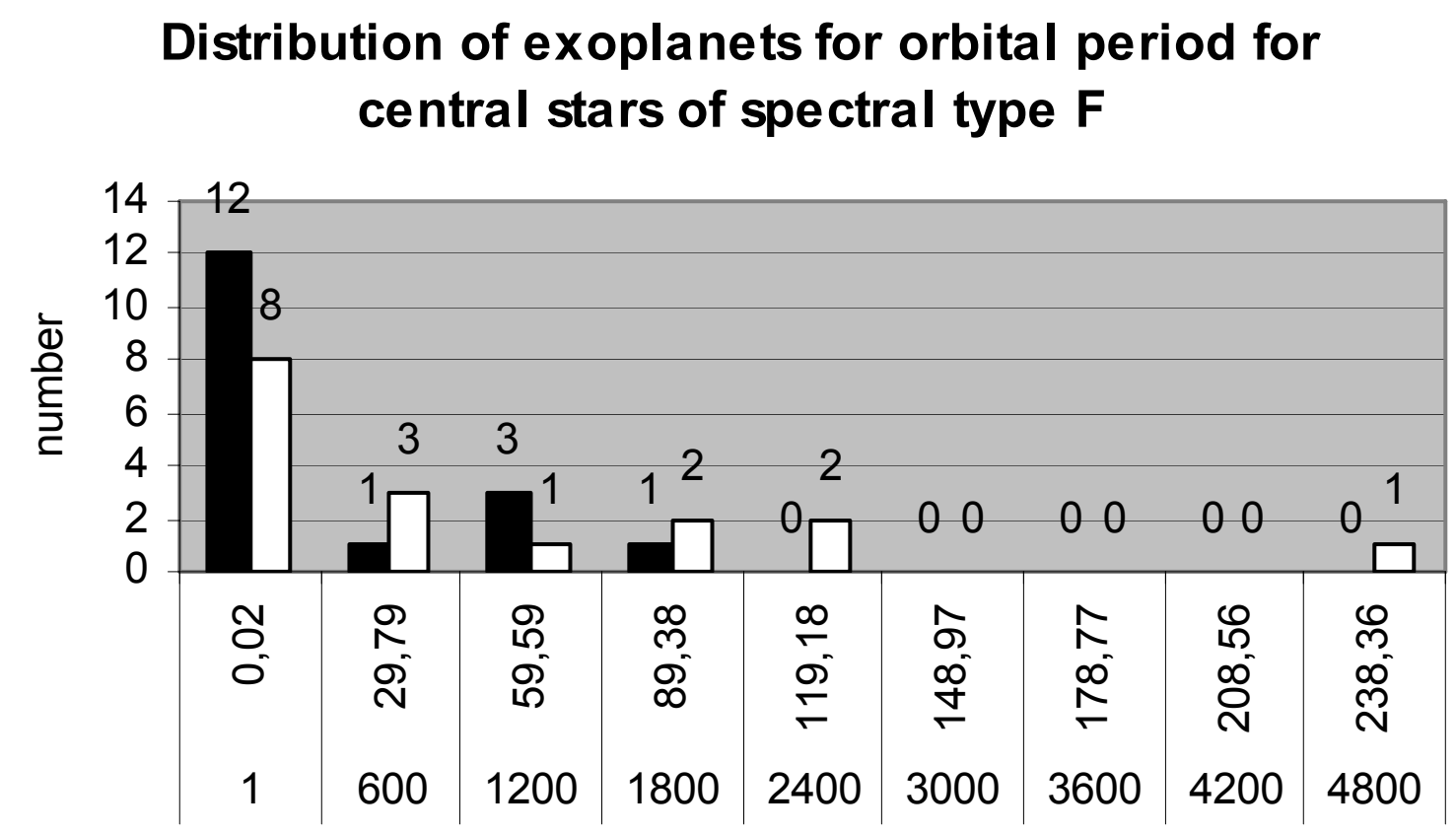

Fig. 12. The histogram distribution of extrasolar planets for the orbital period for stars of the stellar spectral type F. The correlation coefficient values in the self-consistent and not adapted systems of units is equal to 0.941.

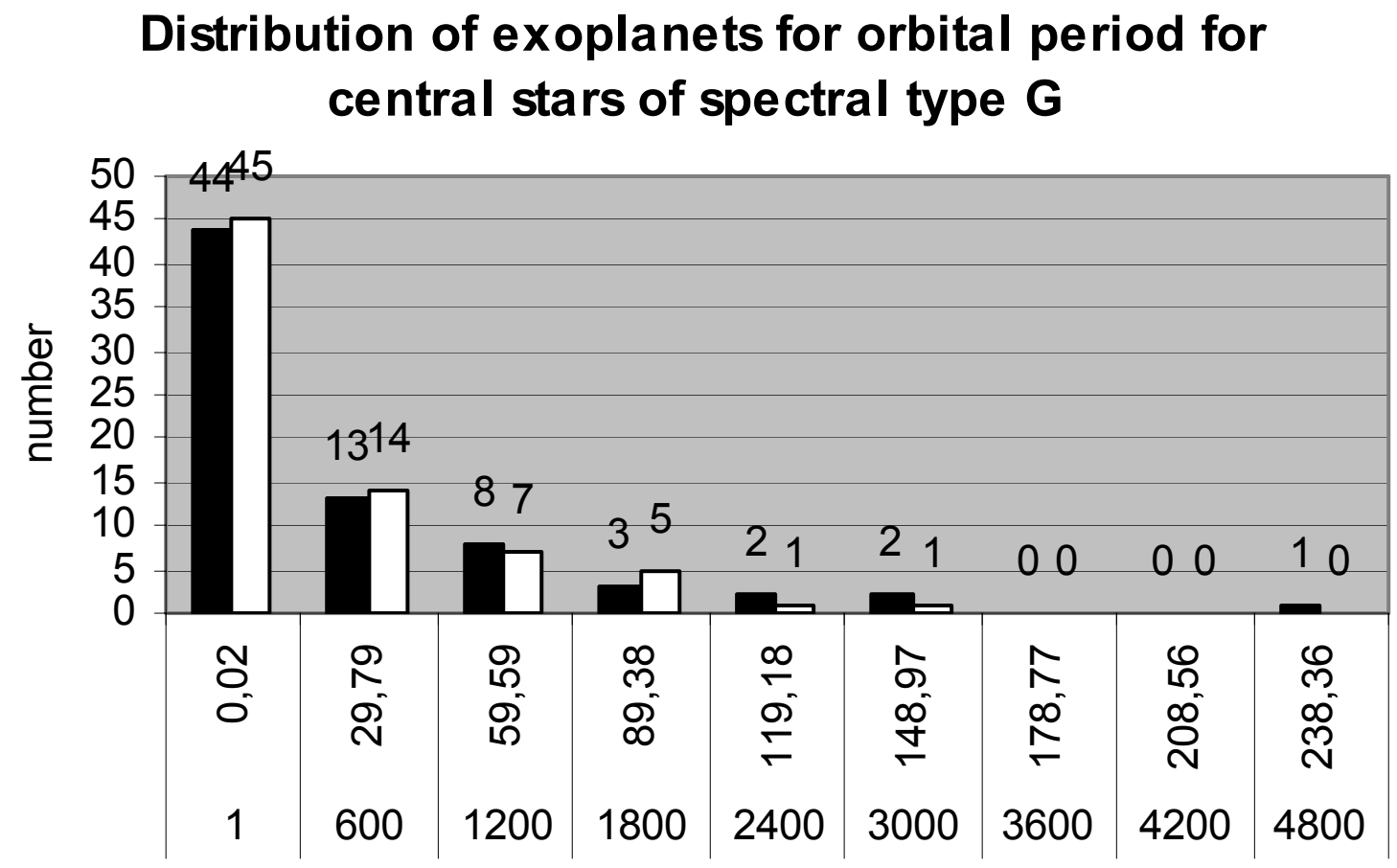

Fig.13. The histogram distribution of extrasolar planets for the orbital period for stars of the stellar spectral type G. The correlation coefficient values in the self-consistent and not adapted systems of units is equal to 0.863.

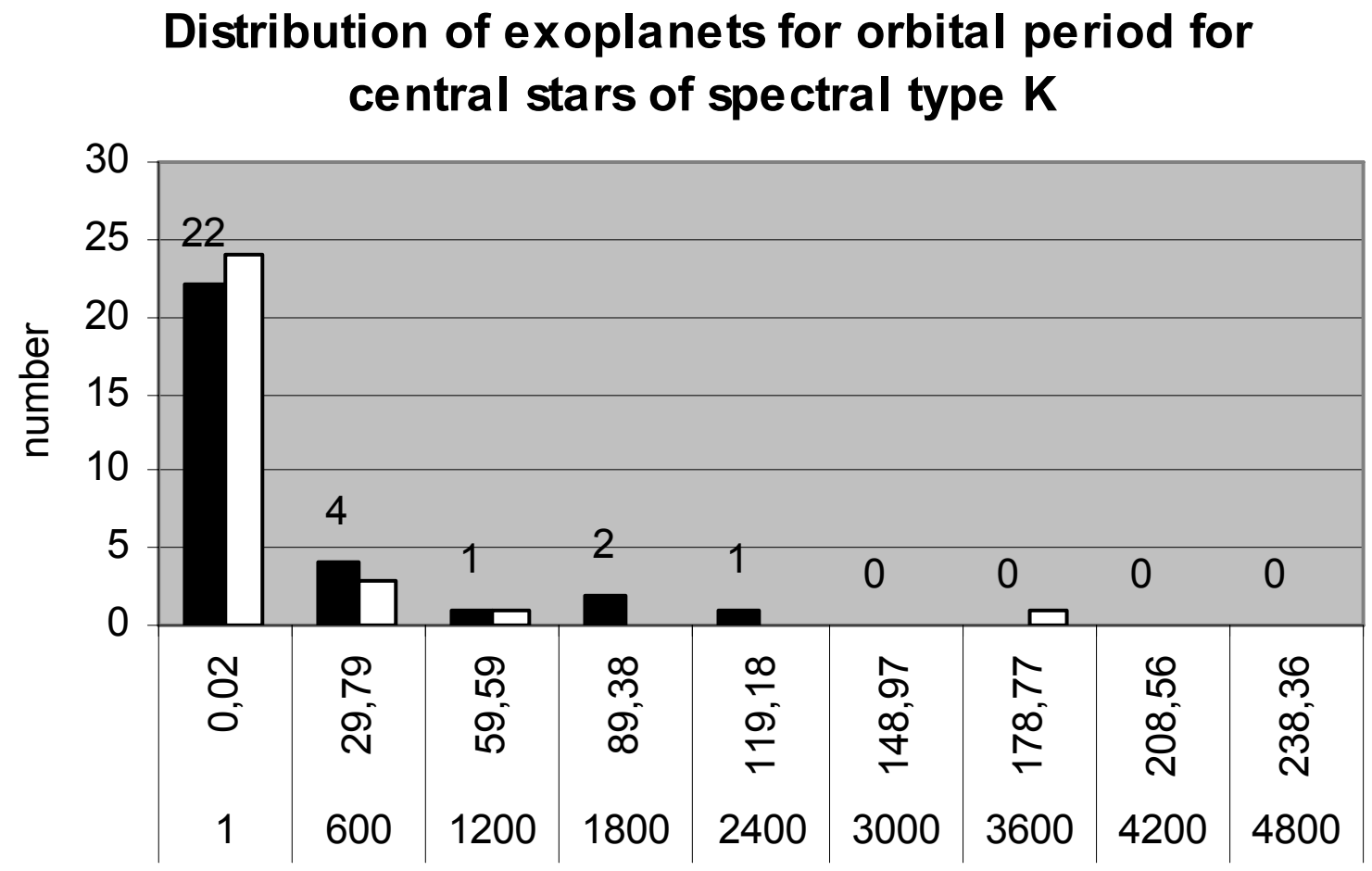

Fig. 14. The histogram distribution of extrasolar planets for the orbital period for stars of the stellar spectral type K. The correlation coefficient values in the self-consistent and not adapted systems of units is equal to 0.883.

From figures 6 and 7 it is visible, that distributions in the self-consistent system of units more monotonous, have obviously expressed maximum.

The common trend in the mass distribution is such that the distribution is not adapted to the system of units, more monotonous.

The distribution of exoplanets for eccentricity appears exponential in nature, which generally persists for homogeneity of the distributions on a sample of individual spectral types of stars, with the exception of spectral types M and F, for which there is insufficient statistical number.

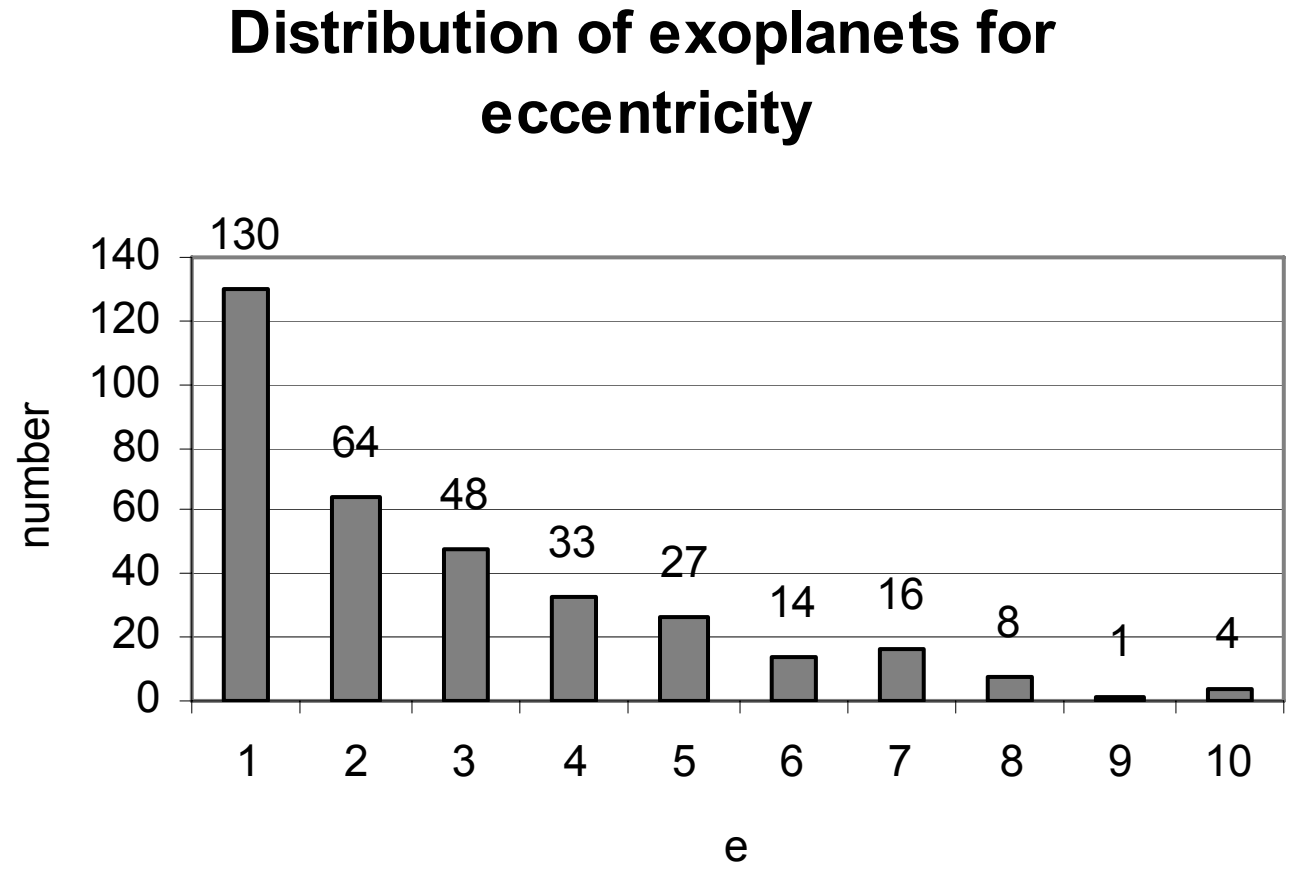

Fig. 15. Histogram of the predicted distribution of the eccentricity of extrasolar planets (in dimensionless units).

**Conclusion**

Given that the dominant role in the dynamical evolution of planetary systems are parameters (and, above all, mass) of the central star, the correct comparison of the dynamic characteristics of exoplanets is possible only in a system of units adapted (self-consistent) to the parameters of the central star. Such approach considers their hierarchy - a dominating role of the central star in evolution of planetary system within the limits of *n*-bodies problem, instead of all statistical sample of planets. While use of not self-consistent parameters leads to distortion of results and does not allow correctly identify the general laws of evolution of exoplanets systems. As consequence, in distributions in not adapted system of units the harmonics connected with not considered dynamic influence of the central stars are shown.

Distributions in the reduced (self-consistent) system of units are exponential. The precise distribution of extrasolar planets by the masses, as well as on the orbital period suggests expressly selective effect.

For each statistical distribution the correlation coefficient between corresponding values of parameters in the self-consistent and not adapted systems of units has been calculated. Divergences between the values of the same name in different systems of units are substantial, and the homogeneity of the distributions is broken when the correlation coefficient smaller than 0.97. As expected, the correlation coefficients of samples for the spectral types of stars are much higher than for the entire sample.

Presence of the common laws proves to be true existence of resonant systems, and, in particular, two-frequency resonances. From 44 multiple systems 17 - are connected by two-frequency orbital resonances of lower orders.

Revealing of more thin effects, including with use “wavelet methods” (*Baluyev R.V.* 2006.), in obviously incorrectly reduced distributions, certainly, is not justified, to what the homogeneous statistical distributions of dynamic parameters of exoplanets received in the present work testify.

**Referents**